\documentclass[12pt, border=55mm,preprintnumbers]{article} 

\usepackage{graphicx}
\usepackage{url}
\usepackage{multirow}
\usepackage{epsfig}
\usepackage{amsmath}
\usepackage{bm}
\usepackage{times}
\usepackage{graphicx}
\usepackage{color}
\usepackage{slashed}
\usepackage{graphicx}
\usepackage{amsmath}
\usepackage[latin1]{inputenc}
\usepackage{hyperref}
\usepackage{soul}
\usepackage{lipsum}
\usepackage{amssymb}
\usepackage{amsmath}

\usepackage{graphicx}
\usepackage{color} 
\usepackage{relsize} 
\usepackage{lmodern}

\usepackage{hyperref}
\usepackage{enumitem}

\usepackage{float}

\def\bea{\begin{eqnarray}}
\def\eea{\end{eqnarray}}
\def\bean{\begin{equation*}}
\def\eean{\end{equation*}} 
\def\nn{\nonumber}
\def\beaal{\begin{align}}
\def\eeaal{\end{align}}

\begin{document}

\thispagestyle{empty}

\noindent\
\\
\\
\\
\begin{center}
\large \bf Neutron Dark Decay\,\footnote{Invited review for the journal {\bf{Universe}} based on:\\ B.\,Fornal and B.\,Grinstein, \emph{Dark Matter Interpretation of the Neutron Decay Anomaly},  Phys.\,Rev.\,Lett.\,120, 191801 (2018) \cite{Fornal:2018eol}.}
\end{center}
\hfill
 \vspace*{0.3cm}
\noindent
\begin{center}
{\bf Bartosz Fornal}\\ \vspace{2mm}
{\emph{Department of Chemistry and Physics, Barry University,\\ 
Miami Shores, Florida 33161, USA}}
\vspace*{1.5cm}
\end{center}

\begin{abstract}
There exists a puzzling disagreement between the results for the neutron lifetime  obtained in experiments using the beam technique  versus those relying on the bottle method. A possible  explanation of this discrepancy postulates the existence of a beyond-Standard-Model decay channel of the neutron involving new  particles in the final state, some of which can be  dark matter candidates. We review the current theoretical status of this proposal and discuss the particle physics models accommodating such a dark decay. We then elaborate on the  efforts undertaken  to test this hypothesis,  summarizing the prospects for probing neutron dark decay channels in future experiments.
\end{abstract}

\newpage

\section{Introduction}	\label{s1}

Almost all of the visible  matter in the Universe is made up of atoms, which consist of electrons, protons, and neutrons. This picture of the subatomic world emerged only less than a century ago. The electron was discovered by Joseph Thomson in 1897 \cite{doi:10.1080/14786449708621070}, whereas the proton was proposed by  Ernest Rutherford in 1919  \cite{Rutherford:1919fnt}. The first hints of the existence of a neutron were provided  by Ir\`{e}ne Joliot-Curie and Fr\'{e}d\'{e}ric Joliot-Curie in early 1932 \cite{Irene}, but the actual  proposal supported by further experiments was put forward by James Chadwick later in 1932 \cite{1932RSPSA.136..692C}. Soon afterwards, Ir\`{e}ne and Fr\'{e}d\'{e}ric Joliot-Curie precisely determined the mass of the neutron \cite{Irene2}, identifying it to be heavier than the proton. This mass relation allows the neutron to decay and has profound implications for the physics at the nuclear level.

\begin{figure}[t!]
\centerline{\includegraphics[trim={0cm 9cm 0cm 1.0cm},clip,width=10cm]{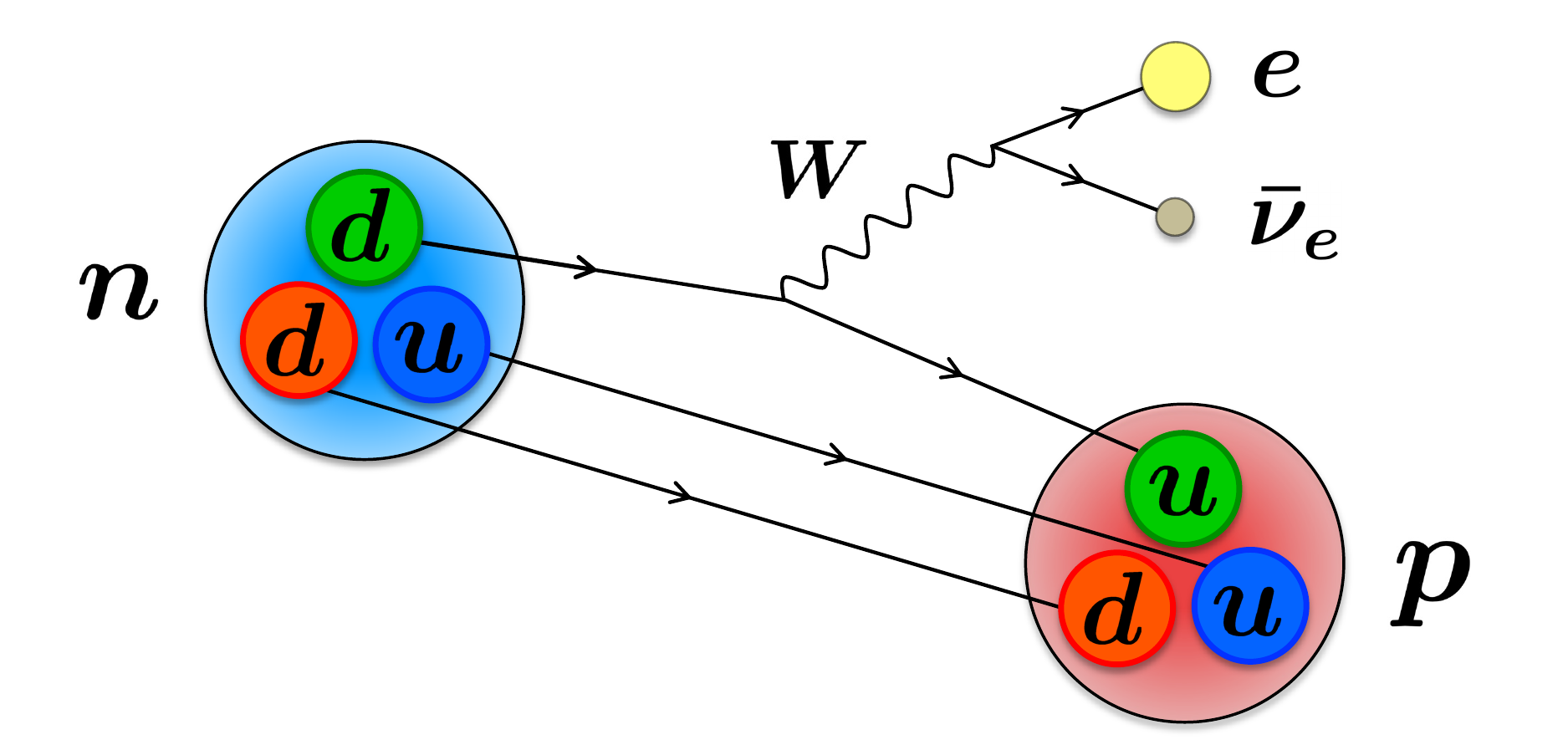}}
\vspace*{0pt}
\caption{\small{Dominant neutron decay  $n \to p + e + \bar\nu_e$ in the Standard Model.}\protect\label{fig:1}}
\vspace{8mm}
\end{figure}

Due to nonzero binding energy inside a nucleus,  the neutron does not decay in stable nuclei. However, on its own it  undergoes a $\beta$ decay predominantly to a proton, electron, and electron antineutrino (see, Figure \ref{fig:1}),
\bea
n \to p + e + \bar\nu_e \ .
\eea 
Apart from this leading order process, there are channels involving photons in the final state, with a branching ratio  ${{\rm Br}(n\to p +e +\bar{\nu}_e+ \gamma)} \sim 1\%$ \cite{Bales:2016iyh}. Finally, there exists also a decay channel to hydrogen and an antineutrino, but the corresponding branching ratio is tiny,  ${{\rm Br}(n\to {\rm H} + \bar{\nu}_e)} \sim 4 \times 10^{-6}$ \cite{Faber:2009ts}.\break 

\vspace{-4mm}
A precise calculation of the neutron lifetime within the framework of the Standard Model of particle physics \cite{Glashow:1961tr,Higgs:1964pj,Englert:1964et,Weinberg:1967tq,Salam:1968rm,Fritzsch:1973pi,Gross:1973id,Politzer:1973fx}, taking into account  radiative corrections, yields the formula \cite{Marciano:2005ec,Czarnecki:2018okw}
\bea\label{nlife}
\tau_n^{\rm th} = \frac{4908.6(1.9) \ {\rm s}}{(1+3\lambda^2)|V_{ud}|^2} \ ,
\eea
where $\lambda$ is the ratio of the axial-vector current coefficient and the vector current coefficient in the matrix element for the neutron $\beta$ decay,
\bea
\mathcal{M} = \frac{G_F}{\sqrt2} V_{ud}\,g_V\left[\,\bar{p}\,\gamma_\mu n - \lambda \,\bar{p}\,\gamma_5\gamma_\mu n\,\right]\left[\,\bar{e}\, \gamma^\mu (1-\gamma_5)\nu\,\right],
\eea
and 
$V_{ud}$ is the Cabibbo-Kobayashi-Maskawa (CKM) matrix element, which, based on 15 most precisely measured  superallowed transitions \cite{Hardy:2014qxa}, has a dispersion-relation-based weighted average of $|V_{ud} |= 0.97373(11)(9)(27)$ \cite{Workman:2022ynf}.\break
 
 \noindent
Using the up-to-date   average for the ratio of the axial-vector to vector current coefficient, $\lambda_{\rm av}= -1.2754 \pm 0.0013$ \cite{Workman:2022ynf}, the resulting neutron lifetime from Eq.\,(\ref{nlife}) is $\tau_n = 880.5 \pm 2.3 \ {\rm s}$. An independent  nuclear  lattice calculation gives  $\lambda_{\rm lattice} = - 1.271 \pm 0.013$ \cite{Chang:2018uxx}, which corresponds to $\tau_n = 885\pm 15\ {\rm s}$.

Due to difficulties in isolating low-energy neutrons, experiments capable of  directly measuring  the neutron lifetime were not performed until fairly recently. At present, there are two qualitatively different techniques implemented to perform this measurement: the beam method and the bottle method. In the future, a third approach using magnetic storage might also be developed. For a review of those methods, including a detailed timeline of neutron lifetime measurements for each of them, see \cite{Dubbers:2011ns,2011RvMP...83.1173W}.

In beam experiments, a collimated beam of cold neutrons passes through a quasi-Penning trap. The protons from neutron decays are trapped and counted, enabling the determination of the rate of neutron decays involving protons in the final state, $dN_p/dt$. The flux of neutrons in the beam is found by counting $\alpha$ particles and tritium nuclei from the $(n,\alpha)$ reaction in a deposit of $^6{\rm LiF}$ on a thin silicon crystal wafer, and the number of neutrons, $N_n$, from which the trapped protons originated is established. The beam neutron lifetime is then calculated as,
\bea
\tau_n^{\rm beam} = - {N_n}\left({\frac{dN_p}{dt}}\right)^{-1} \!.
\eea

\newpage
In  the bottle  method, ultracold neutrons are stored in a specially prepared container, which is emptied at variable storage times and the neutrons inside it are counted using, e.g., a $^3{\rm He}$ proportional counter, thus determining $N_n$ as a function of time.  Since the decay pattern is exponential, the bottle neutron lifetime is obtained  from the fit  to the experimental points of the function
\bea
N_n(t)=N_0 \exp\!\left(-t/\tau_n^{\rm bottle}\right).
\eea

\begin{figure}[t!]
\centerline{\includegraphics[trim={2cm 1.1cm 2.7cm 0},clip,width=9.9cm]{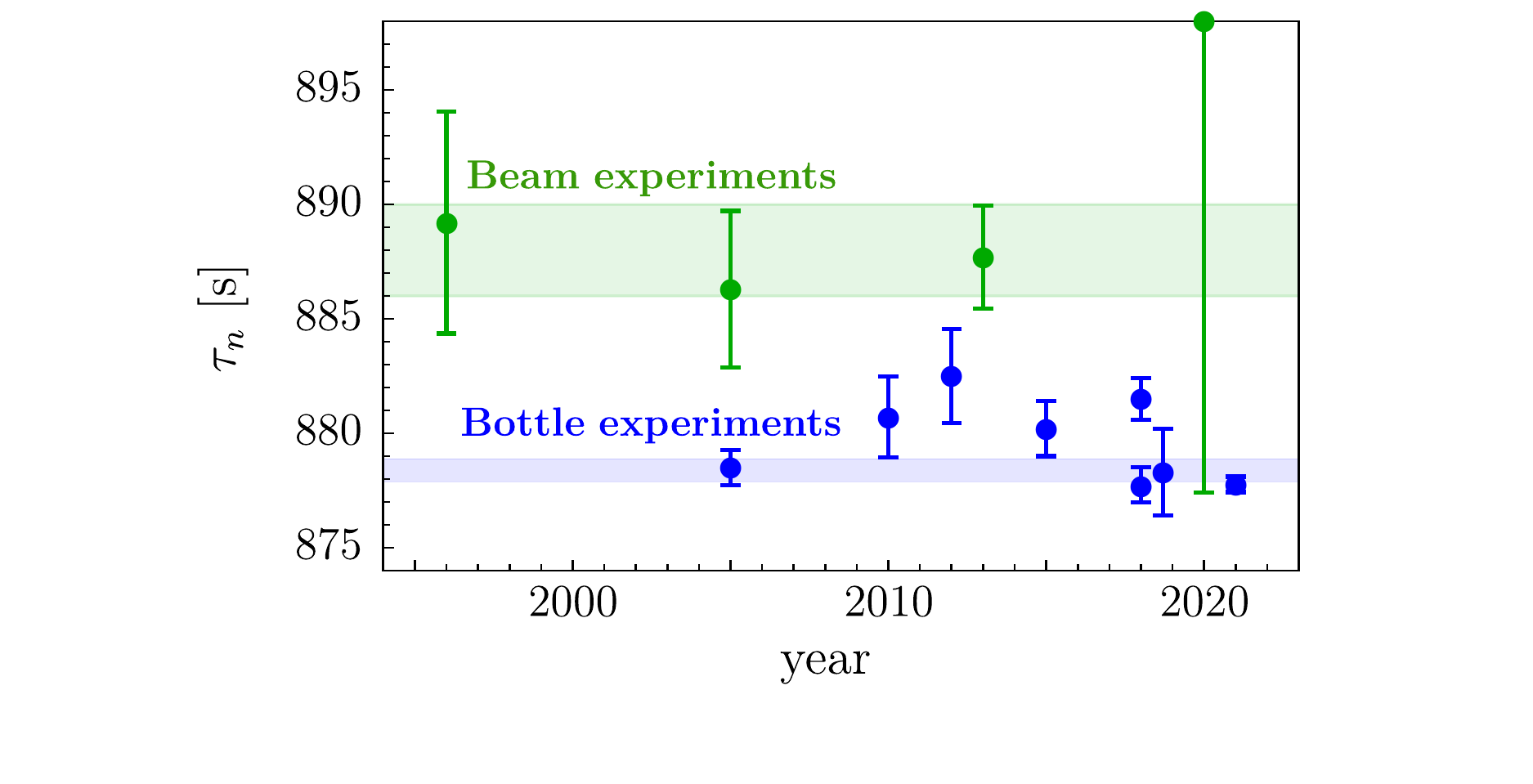}}
\vspace*{-6pt}
\caption{\small{Neutron lifetime measurements in beam experiments (green) \cite{Byrne:1996zz,Nico:2004ie,Yue:2013qrc,Hirota:2020mrd}  and bottle experiments (blue)  \cite{Serebrov:2004zf,Pichlmaier:2010zz,Steyerl:2012zz,Arzumanov:2015tea,Serebrov:2017bzo,Pattie:2017vsj,Ezhov:2014tna,UCNt:2021pcg} made over the last 30 years. The 2013 beam result is based on a reanalysis of the 2005 data.}\protect\label{fig:2}}
\vspace{8mm}
\end{figure}

Figure \ref{fig:2} presents a summary of the recent neutron lifetime measurements, including the available beam experiment results  \cite{Byrne:1996zz,Nico:2004ie,Yue:2013qrc,Hirota:2020mrd} and bottle experiment results  \cite{Serebrov:2004zf,Pichlmaier:2010zz,Steyerl:2012zz,Arzumanov:2015tea,Serebrov:2017bzo,Pattie:2017vsj,Ezhov:2014tna,UCNt:2021pcg}. There is a clear  mismatch between the two types of measurements. The average neutron lifetime from the beam experiments \cite{Byrne:1996zz,Yue:2013qrc} is
\bea\label{beamm}
\tau_n^{\rm beam} = 888.0 \pm 2.0 \ {\rm s} \ ,
\eea
while the average from bottle experiments is
\bea\label{bottleb} 
\tau_n^{\rm bottle} = 878.4\pm 0.5 \ {\rm s}\ .
\eea
The discrepancy is at the level of four standard deviations. Although this might be due to unaccounted for systematic errors,  some sort of new physics affecting  the way neutron decays is a viable possibility to consider.

Since in the Standard Model one always expects a proton in the final state of neutron decay, the beam and bottle experiments should give the same result, with just a tiny difference due to the decay channel  $n\to {\rm H} + \bar{\nu}_e$ (the beam experiment is not sensitive to it). Therefore, within the Standard Model one has ${\rm Br}(n\to p +{\rm anything})_{\rm SM} = 1$ with accuracy $\mathcal{O}(10^{-5})$. 

The crucial observation is that one can reconcile the results in Eqs.\,(\ref{beamm}) and (\ref{bottleb}) if the neutron has one or more extra decay channels for which none of the final state particles is a proton.
Indeed, since the beam method is not sensitive to them, the beam and bottle  lifetimes are related by
\bea
 \tau_n^{\rm bottle} = {\tau_n^{\rm beam}}\times{{\rm Br}(n\to p +{\rm anything})} \le {\tau_n^{\rm beam}} \ ,
\eea
so in the presence of protonless neutron decay channels it is possible to have a relation between the two measurement of the form
$ \tau_n^{\rm bottle}  <  \tau_n^{\rm beam}$, exactly as the current beam and bottle results suggest.

More precisely, if the branching ratio for neutron beta decays is 
\bea\label{99pp}
{{\rm Br}(n\to p +{\rm anything})} \approx 99\% \ , 
\eea
and the remaining channels corresponds to beyond-Standard-Model neutron dark decays  not involving a proton in the final state, 
\bea\label{1ppp}
{{\rm Br}(n \to {\rm anything}  \ne p)} \approx 1 \% \ ,
\eea
the two experimental results remain consistent with each other. This lies at the heart of the idea proposed in \cite{Fornal:2018eol}, where  phenomenologically viable extensions of the Standard Model were constructed with Eqs.\,(\ref{99pp}) and (\ref{1ppp}) satisfied, as summarized below.

The plan for the review is as follows:
In Section \ref{s2} we analyze the neutron dark decay scenario at the  effective field theory level, deriving the general conditions that the corresponding models need to satisfy, and  we discuss the various possible final states. In Section \ref{s3} we construct concrete particle physics models with neutron dark decays occurring at a branching ratio of $1\%$,  which are consistent with all other experiments. In Section \ref{s4} we discuss the progress made on the experimental side to test the neutron dark decay proposal, and in Section \ref{s5} we elaborate on the theoretical developments in this area, including those  connecting neutron dark decay  to other open questions in particle physics. A brief summary and conclusions are presented in Section \ref{s6}.

\section{Effective picture of neutron dark decay}\label{s2}

Here we review the model-independent requirements that theories involving neutron dark decays need to satisfy to be phenomenologically acceptable. We are considering the neutron decaying to two or more particles, at least one of which is a new (dark) particle from outside of the Standard Model, hence the name \emph{neutron dark decay}. In the most interesting scenarios one or several of those dark particles are viable dark matter candidates. 

{\center{
\subsection*{Stability of nuclei}}}\label{stabilityn}

Let us denote the final state of neutron dark decay by $f$ and the sum of final state particle masses by $M_f$. This sum obviously cannot be heavier than the neutron itself, thus $M_f < m_n$. It also cannot be too small, since otherwise stable nuclei would undergo dark decays, contrary to what is observed. The lower bound can be derived through the following simple reasoning.

If in a nucleus with atomic number $Z$ and mass number $A$ one of its neutrons underwent a dark decay, this would lead to  $(Z,A) \to (Z,A-1)^* + f$. The final state excited nucleus $(Z,A-1)^*$ would then de-excite emitting secondary particles, mostly photons. However, such signatures were looked for at the  Sudbury Neutrino Observatory  \cite{Ahmed:2003sy} and at the Kamioka Liquid Scintillator Antineutrino Detector \cite{Araki:2005jt} and none were discovered. As a result, those searches placed a lower bound on the neutron lifetime from such generic \emph{nuclear dark decays} of $\tau_n \gtrsim 6\times 10^{29} \ {\rm years}$, obviously ruling out neutron dark decays with a branching ratio of $1\%$.

However, there exists a nonzero parameter space for which nuclear dark decays are forbidden despite neutron on its own still undergoing dark decays. The mechanism behind this is identical to the standard one which prevents stable nuclei from decaying despite the neutron itself undergoing $\beta$ decay.
Indeed, if the following mass relation holds,
\bea\label{cons}
m_n-S_n< M_f<m_n \ ,
\eea
where $S_n$ is the separation energy for a neutron  in the nucleus, then the nuclear dark decay $(Z,A) \to (Z,A-1)^* + f$ is 
kinematically forbidden, while the neutron can still undergo a dark decay when by itself. The most stringent constraint in Eq.\,(\ref{cons}) is provided  by the stable nucleus which has the lowest neutron separation energy -- this happens to be  beryllium-9 with $S_n(^9{\rm Be}) = 1.664 \ {\rm MeV}$. Actually, a slightly stronger bound applies in this case, since a $^9{\rm Be}$ decay would proceed via $^9{\rm Be} \to \, ^8{\rm Be}^* + f \to 2 \alpha + f$ due to a rapid disintegration of excited beryllium-8 to two $\alpha$ particles, lowering the threshold for the nuclear dark decay by  $\sim 93 \ {\rm keV}$ \cite{Pfutzner:2018ieu}. Taking this into account, the constraint on the neutron dark decay final state mass is
\bea\label{const1}
937.993 \ {\rm MeV} < M_f < 939.565 \ {\rm MeV} \ .
\eea
This relation assures also the stability of the proton, which would otherwise undergo dark decays via $p \to f + e^+ +\nu_e$ if $M_f < m_p - m_e = 937.761 \ {\rm MeV}$.

Intriguingly, there exist unstable nuclei whose decay pattern might be affected by neutron dark decays, since those decays are allowed if the neutron separation energy for  unstable nuclei is $S_n < 1.572 \ {\rm MeV}$. Searches for such nuclear dark decays will be discussed in Section \ref{s4}.

{\center{
\subsection*{Neutron dark decay channels}}}

The fact that neutron dark decays are phenomenologically viable when the relation in Eq.\,(\ref{const1}) holds, opened the gates to entirely new model-building opportunities not considered before. As mentioned earlier, the new neutron decay channels involve at least one beyond-Standard-Model particle in the final state. Those new particles can be either fermions (denoted  by $\chi$) or bosons (scalars denoted by $\phi$ and vectors denoted by $V$). The simplest possible dark decay channels for the neutron include:
\bea
n \to \chi \, \gamma \ , \ \ \ \ n \to \chi\,\phi \ ,  \ \ \ ... \ \ \ ,
\eea
where the decays not written out explicitly can include other dark particles, photons, neutrinos, and electrons/positrons in the final state.

In an effective Lagrangian picture  of neutron dark decays, the process is governed by the mixing term between the neutron and the dark fermion $\chi'$,
\bea\label{effLag}
\mathcal{L}^{\rm eff}_{\rm mix} =  \varepsilon \,(\bar{n}\,\chi'+\bar\chi' n) \ ,
\eea
where $\varepsilon$ is a model-dependent parameter with dimension of mass. This dark fermion $\chi'$ can either be the final state particle $\chi$ from the decay $n \to \chi \, \gamma$, or an intermediate fermion $\tilde\chi$ allowing for  the decay $n \to \chi\,\phi$. Both of those cases are discussed in detail below. 

\newpage

{\center{
\subsection*{Neutron $\,\to\,$ dark fermion $+$ photon}}}

In the effective field theory picture the case $n \to \chi \, \gamma$ requires only one new fermion $\chi$. Its mass is constrained by the condition in Eq.\,(\ref{const1}) to satisfy
\bea\label{range1}
937.993 \ {\rm MeV} < m_\chi < 939.565 \ {\rm MeV} \ ,
\eea
thus the energy of the monochromatic photon in the final state, depending on the mass of $\chi$, takes a value within the range
\bea
0 < E_\gamma < 1.572 \ {\rm MeV} \ .
\eea
In the limit $m_\chi \to m_n$, the energy of the monochromatic photon  $E_\gamma \to 0$.

The final state fermion $\chi$, if stable, may be a good dark matter candidate. 
In that case, to prevent $\chi$ from $\beta$ decaying one requires
\bea\label{range111}
937.993 \ {\rm MeV} < m_\chi < m_p+m_e  = 938.783 \ {\rm MeV} \ ,
\eea
which considerably narrows down the range of the expected monochromatic photon energies to 
\bea\label{mprange}
0.782 \ {\rm MeV} < E_\gamma < 1.572 \ {\rm MeV} \ .
\eea
This very concrete prediction was the trigger for experimentalists to start looking for such a signal  immediately after the proposal in \cite{Fornal:2018eol} was made \cite{Tang:2018eln}.

A quantitative description of the decay channel $n \to \chi \, \gamma$ is obtained by constructing an effective Lagrangian containing the terms in Eq.\,(\ref{effLag}) and the neutron's magnetic moment interaction, 
 \bea\label{effLagfull}
\mathcal{L}_{I}^{\rm eff} =\bar{n}\,\Big(i\slashed\partial-m_n +\frac{g_ne}{8 m_n}\sigma^{\,\mu\nu}F_{\mu\nu}\Big) \,n
+  \bar{\chi}\,\big(i\slashed\partial-m_\chi\big) \,\chi + \varepsilon \left(\bar{n}\,\chi + \bar{\chi}\,n\right)  , \ 
\eea
where $g_n$ is the  $g$-factor of the neutron. The rate for the neutron dark decay  $n \to \chi \, \gamma$  is given by
\bea\label{ratedark1}
\Delta\Gamma_{n\rightarrow \chi\gamma} = \frac{g_n^2e^2}{128\pi}\bigg(1-\frac{m_\chi^2}{m_n^2}\bigg)^3  \left(\frac{\varepsilon}{m_n\!-\!m_\chi}\right)^2 m_n  \ .
\eea
If $\Delta\Gamma_{n\rightarrow \chi\gamma}/ \Gamma_{n} \approx 1\%$, where $\Gamma_{n}$ is the total neutron decay rate, this provides a viable explanation for the observed discrepancy between the beam and bottle experiments. The corresponding particle physics framework for this case (Model 1) will be constructed in Section \ref{s3}.

\newpage

{\center{
\subsection*{Neutron $\,\to\,$ dark fermion $+$ dark scalar}}}

The effective theory for the pure dark decay $n\to \chi\,\phi$ contains two new fermions $\chi$ and $\tilde\chi$, as well as a new scalar $\phi$. In general, the scalar $\phi$ can be replaced by a vector boson $V_\mu$. The dark fermion $\tilde\chi$ is an intermediate particle mixing with the neutron and coupling to $\chi$ and $\phi$. 
 The sum of masses $m_\chi+m_\phi$ is constrained by the condition in Eq.\,(\ref{const1}) to satisfy
\bea\label{now23}
937.993 \ {\rm MeV} < m_\chi + m_\phi< 939.565 \ {\rm MeV} \ ,
\eea
whereas the mass of  $\tilde\chi$ is bounded from below, $m_{\tilde\chi} > 937.993 \ {\rm MeV}$, to prevent the decay of $^9{\rm Be}$ triggered by the neutron dark decay $n\to \tilde\chi\,\gamma$. 

The final state fermion $\chi$ and scalar $\phi$ can be dark matter candidates if they are stable, which happens when 
\bea
|m_\chi-m_\phi|< m_p+m_e = 938.783 \ {\rm MeV} \ .
\eea
Apart from the relations above, there are no other constraints on the masses $m_\chi$ and $m_\phi$, e.g., one can have  $m_\chi \gg m_\phi$, $m_\chi \ll m_\phi$, or
$m_\chi \approx m_\phi $.

\vspace{1mm}
The case ${n\to \chi\phi}$ can be effectively described by the Lagrangian
 \bea\label{lageff1122}
\mathcal{L}_{I\!I}^{\rm eff}\!\!&=&\!\!\mathcal{L}_{I}^{\rm eff}\left(\chi \to \tilde\chi \right)+(\lambda_\phi \,\bar{\tilde\chi}\,\chi \,\phi + {\rm h.c.}) \nn\\
\!\!&+& \!\! \bar{\chi}\,\big(i\slashed\partial-m_\chi\big) \,\chi \ + \partial_\mu \phi^*\partial^\mu \phi - m_\phi^2|\phi|^2 \ ,
\eea
which results in the neutron dark decay rate
\bea\label{rateT22}
\Delta\Gamma_{n\rightarrow \chi\phi} = \frac{|\lambda_\phi|^2}{16\pi} \,\sqrt{\left[(1+x)^2-y^2\right]\left[(1-x)^2-y^2\right]^3}  \left(\frac{\varepsilon}{m_n\!-\!m_{\tilde\chi}}\right)^2 m_n  \ ,
\eea
where $x= m_\chi/m_n$ and $y=m_\phi/m_n$. If $m_{\tilde\chi} > m_n$ then ${n\to \chi\,\phi}$ is the only dark decay channel available and provides a solution to the neutron lifetime  discrepancy if $\Delta\Gamma_{n\rightarrow \chi\phi} / \Gamma_{n} \approx  1\%$. However, if $m_n>m_{\tilde\chi} > 937.993 \ {\rm MeV}$, then an additional dark decay channel opens up, $n\to \tilde\chi\,\gamma$, with a decay rate
$
\Delta\Gamma_{n\rightarrow \tilde\chi\gamma} = \Delta\Gamma_{n\rightarrow \chi\gamma} \left(m_\chi \to m_{\tilde\chi}\right).
$
If the  following relation is satisfied,
\bea
\left(\Delta\Gamma_{n\rightarrow \chi\phi}+\Delta\Gamma_{n\rightarrow \tilde\chi\gamma}\right)\!/ \Gamma_{n}  \approx  1\% \ , 
\eea
this also explains the discrepancy between the beam and bottle experiments. The corresponding particle physics model in this case is constructed below (Model 2).  We  note that this picture can be further simplified if the fermion $\chi$ also plays the role of the intermediate particle $\tilde\chi$.

\newpage

\section{Particle physics models}\label{ppmodels}
\label{s3}

In this section we discuss the two simplest microscopic renormalizable models providing realizations  of the neutron dark decay cases $n\to \chi\,\gamma$ and $n\to \chi\,\phi$. Those models were constructed in  \cite{Fornal:2018eol}, but it was later shown that consistency with the observed neutron star masses requires additional interactions to be present, which we will review in Section {\ref{s5}}.

{\center{
\subsection*{Model 1}}}\label{model1o}

The minimum number of new fields needed at the particle physics level to realize the neutron dark decay $n\to \chi\,\gamma$ is two: the Standard Model singlet Dirac fermion $\chi$ in the final state discussed earlier, and a color triplet  scalar $\Phi = (3,1)_{-1/3}$ producing  the mixing terms in Eq.\,(\ref{effLag}) between the neutron and $\chi$. The Lagrangian for such a theory includes the interactions, 
\bea\label{lagMod1}
\mathcal{L}_1 \ \supset\  \lambda_q \epsilon^{ijk}\overline{u^c_{L}}_i d_{Rj}\Phi_k+ \lambda_\chi \Phi^{*i}\bar{\chi}\,d_{Ri} \ + \ {\rm h.c.} \ ,
\eea
where the superscript $c$  denotes charge conjugation. It is possible to define a conserved generalized baryon number $B$ if one assigns $B_\chi=1$, $B_\Phi=-2/3$, and the standard $B_q = 1/3$  to the quarks.

\begin{figure}[t!]
\centerline{\includegraphics[trim={0cm 13cm 2cm 1cm},clip,width=10cm]{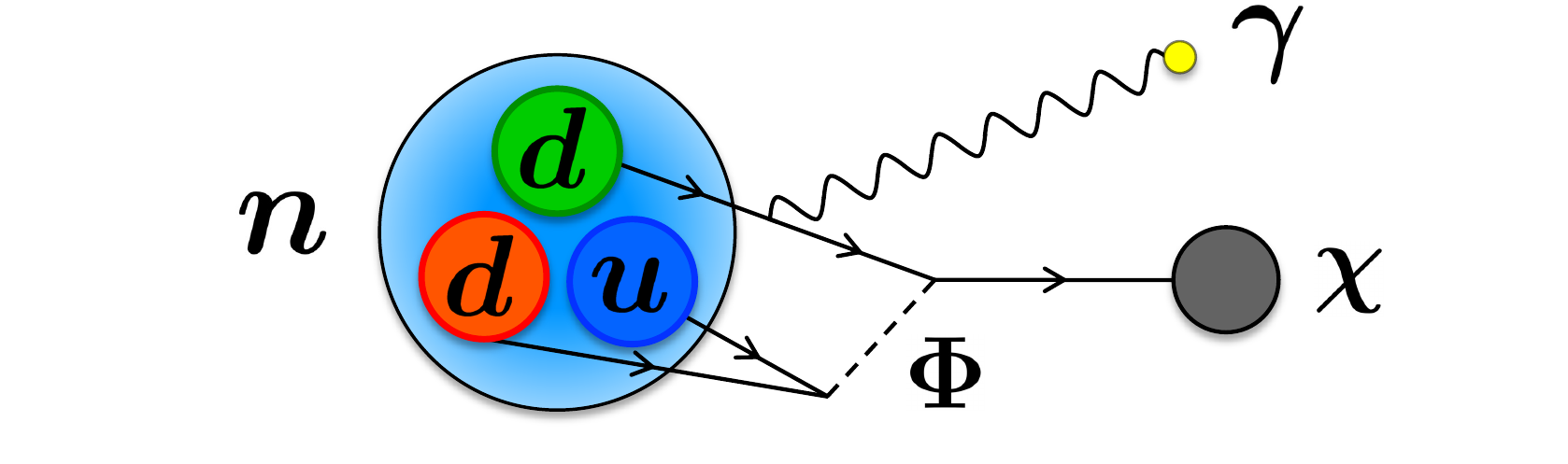}}
\vspace*{2pt}
\caption{\small{Neutron dark decay $n\to \chi\,\gamma$ in Model 1.}\protect\label{fig:3}}
\vspace{6mm}
\end{figure}

Figure \ref{fig:3} shows a schematic diagram of the neutron dark decay  $n\to \chi\,\gamma$ in Model 1. Upon matching the effective Lagrangian  in Eq.\,(\ref{effLagfull}) with the particle-level Lagrangian in Eq.\,(\ref{lagMod1}), the decay rate  is given by  Eq.\,(\ref{ratedark1}) with
\bea
\varepsilon = \beta\, \frac{\lambda_q\lambda_\chi}{m_\Phi^2}\ ,
\eea
where $\beta = 0.0144(3)(21) \ {\rm GeV}^3$ is determined from a lattice calculation \cite{Aoki:2017puj}.

The required neutron dark decay rate of $\Delta\Gamma_{n\rightarrow \chi\gamma}/\Gamma_{n}\approx  1\%$ is achieved for phenomenologically viable values of model parameters. In particular, setting $m_\chi \approx 938 \ {\rm MeV}$, i.e., at its minimal value given the requirement in Eq.\,(\ref{range1}), the particle physics parameter choice leading to the $1\%$ branching ratio is
\bea\label{b5}
{m_\Phi} \approx (200 \ {\rm TeV}) {\sqrt{|\lambda_q\lambda_\chi|}} \ .
\eea 
All collider bounds are satisfied for  $m_\Phi \gtrsim 1 \ {\rm TeV}$, while the dinucleon decay \cite{Gustafson:2015qyo} and neutron-antineutron oscillation \cite{Abe:2011ky} constraints do not apply since 
$\chi$ is a Dirac fermion with nonzero baryon number.

{\center{
\subsection*{Model 2}}}\label{model2o}

For the decay $n\rightarrow \chi\phi$ to take place, four new fields at the particle physics level are introduced: Standard Model singlet Dirac fermions $\chi$,  $\tilde\chi$, scalar $\phi$, and the same color triplet scalar $\Phi = (3,1)_{-1/3}$ as in Model 1. The corresponding Lagrangian contains the terms
\bea\label{lagMod2}
\mathcal{L}_2 \ \supset \ \lambda_q \epsilon^{ijk}\overline{u^c_{L}}_i d_{Rj}\Phi_k+ \lambda_{\tilde\chi} \Phi^{*i}\bar{\tilde\chi}\,d_{Ri} 
+ \lambda_\phi\bar{\tilde\chi}\,\chi\,\phi \ + \ {\rm h.c.}  \ .
\eea
Baryon number is again conserved if $B_\phi=0$, $B_{\tilde{\chi}}=B_\chi = 1$, and $B_\Phi=-2/3$.

\begin{figure}[t!]
\centerline{\includegraphics[trim={0cm 13cm 2cm 1.3cm},clip,width=10cm]{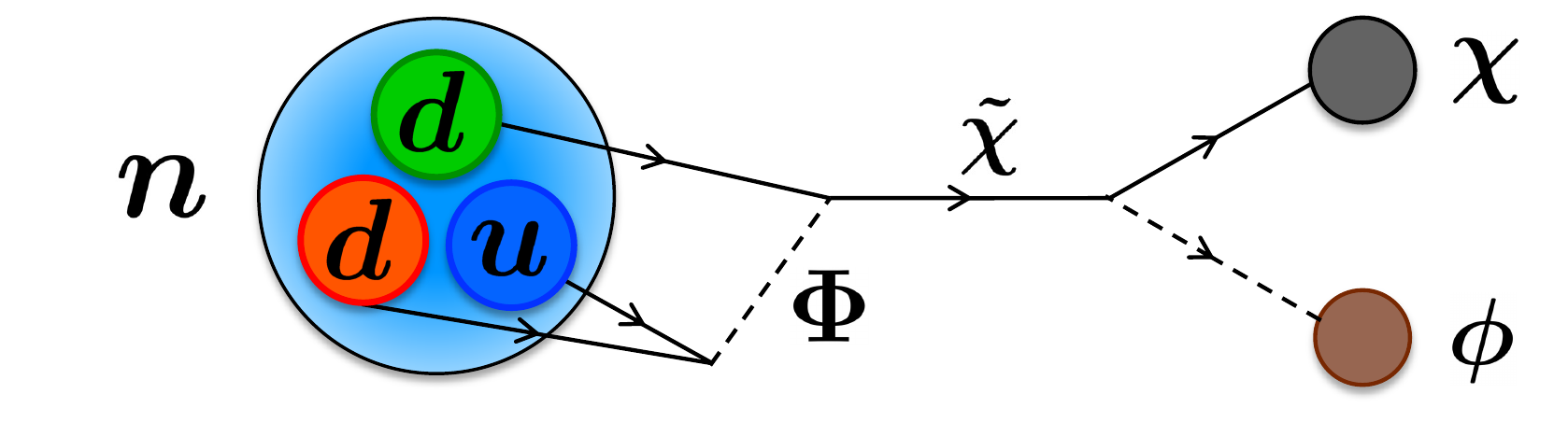}}
\vspace*{2pt}
\caption{\small{Neutron dark decay $n\to \chi\,\phi$ in Model 2.}\protect\label{fig:4}}
\vspace{4mm}
\end{figure}

Figure \ref{fig:4} shows  a schematic diagram of the neutron dark decay  $n\to \chi\,\phi$ in Model 2. Comparing the Lagrangians in Eqs.\,(\ref{lageff1122}) and (\ref{lagMod2}), the matching procedure yields the decay rate given by Eq.\,(\ref{rateT22}) with  $\varepsilon = {\beta\,\lambda_q\lambda_{\tilde\chi}}/{m_\Phi^2}$. The required decay rate of $\Delta\Gamma_{n\rightarrow \chi\phi} / \Gamma_{n}\approx 1\%$ is achieved for many parameter choices, e.g., $m_\chi = 938 \ {\rm MeV}$, $m_\phi \ll m_\chi$, $m_{\tilde\chi} = 2 \,m_n$, and
\bea\label{b2}
{m_\Phi}} \approx (300 \ {\rm TeV}){\sqrt{|\lambda_q\lambda_{\tilde\chi}\lambda_\phi|} \ ,
\eea 
again remaining consistent with experimental constraints. To simplify the model even further, the intermediate fermion $\tilde\chi$ can be taken to be the same $\chi$ particle that appears in the final state of neutron dark decay.

\newpage

\section{Experimental progress}\label{s4}

Searches for signatures of neutron dark decay were initiated right after the neutron dark decay proposal in \cite{Fornal:2018eol} was put forward. They involved looking  for the  photon from neutron dark decays $n\to\chi\,\gamma$, the electron-positron pair from $n\to \chi\,e^+e^-$, and nuclear dark decay signatures triggered by $n\to \chi\,\phi$. In addition, several new search strategies have been proposed to look for signs of neutron dark decay in other ongoing and future experiments.

{\center{
\subsection*{Search for $\boldsymbol{n\to \chi\,\gamma}$}}}

A dedicated experiment searching for the monochromatic photon from the neutron dark decay $n\to\chi\,\gamma$ was carried out at the Ultracold Neutron (UCN) facility at Los Alamos \cite{Tang:2018eln}  within a few weeks after the idea in \cite{Fornal:2018eol} was proposed. It was sensitive to photon energies $0.782 \ {\rm MeV} < E_\gamma < 1.664 \ {\rm MeV}$, thus covering the entire range  expected if $\chi$ is  a dark matter candidate. The results were negative and excluded a dark decay branching ratio of ${\rm Br}(n\to \chi\,\gamma) = 1\%$ at a significance of $2.2$ standard deviations.

Further analysis of this data  was carried out in  \cite{McKeen:2020zni}, where more detailed bounds on the  $n\to\chi\,\gamma$ decay channel were derived as a function of the $\chi$ mass (see, Figure \ref{fig:5}). The green-shaded region corresponds to the parameter space ruled out by the UCN experiment \cite{Tang:2018eln} at $90\%$ confidence level, whereas the purple-shaded  region is excluded by the Borexino experiment \cite{Agostini:2015oze}. It is clear that the neutron dark decay with a branching ratio of $1\%$, denoted by the black line, is in tension with those results. It is worth pointing out that in the mass range $m_\chi \gtrsim 938.470 \ {\rm MeV}$, i.e., for the monochromatic photon energies $E_\gamma \lesssim 1.095 \ {\rm MeV}$, the neutron dark decay at the level ${\rm Br}(n\to \chi\,\gamma) \lesssim 0.1\%$ is not constrained by those experiments.

\begin{figure}[t!]
\centerline{\includegraphics[trim={2cm 0.5cm 2.5cm 0.0cm},clip,width=9cm]{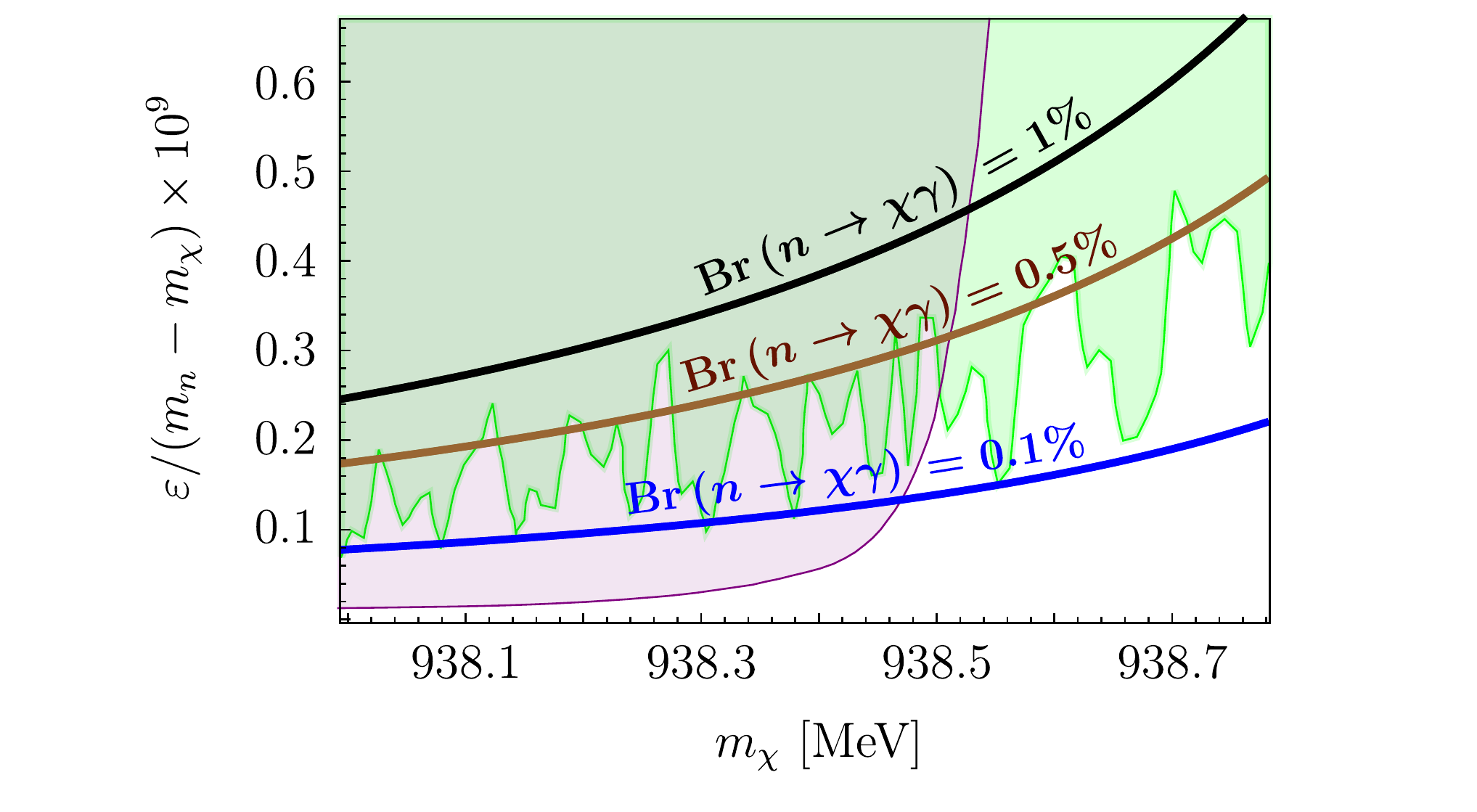}}
\vspace*{-4pt}
\caption{\small{Plot of the parameter space of  dark matter mass $m_\chi$ versus the mixing $\varepsilon/(m_n-m_\chi)$ with the curves corresponding to the branching ratios for $n\to\chi\,\gamma$ of $1\%$ (black), $0.5\%$ (brown), and $0.1\%$ (blue). The green-shaded region denotes the $90\%$ confidence level exclusion from the Los Alamos UCN experiment \cite{Tang:2018eln}, whereas the purple-shaded region corresponds to the exclusion from Borexino \cite{McKeen:2020zni,Agostini:2015oze}.}\label{fig:5}}
\vspace{5mm}
\end{figure}

{\center{
\subsection*{Search for $n\to \chi\,e^+e^-$}}}

One month later, an analysis of another set of  Los Alamos UCN data was performed in the search of the electron-positron pairs from the possible neutron dark decay channel $n\to \chi\,e^+e^-$ \cite{Sun:2018yaw}. However, no $e^+e^-$ pairs were found, and a stringent bound of  $(E_{ee} -2\,m_e) \gtrsim 100 \ {\rm keV}$ was derived for the branching ratio ${\rm Br}(n\to \chi\,e^+e^-) = 1\%$. This constraint was improved the following year by the PERKEO II experiment \cite{Klopf:2019afh}, and a strong exclusion was set for the energy region  $(E_{ee} -2\,m_e)\gtrsim 30 \ {\rm keV}$.

{\center{
\subsection*{Search for nuclear dark decays}}}\label{ndd}

As was shown in Section \ref{s2}, all stable nuclei are safe from dark decays if the final state of neutron dark decay has a mass $M_f > 937.993 \ {\rm MeV}$. This bound is saturated by $^9{\rm Be}$, which has the lowest neutron separation energy, $S_n(^{9}{\rm Be}) = 1.664 \ {\rm MeV}$, out of all stable nuclei. Nevertheless, many unstable nuclei have  neutron separation energies $S_n$ smaller than that of $^9{\rm Be}$, which makes dark decays of such nuclei possible if the $\chi$ and $\phi$ particles are sufficiently light, i.e., if the following mass relation holds,
\bea
937.993 \ {\rm MeV} < M_f < m_n - S_n \ .
\eea

An example of such an unstable nucleus is $^{11}{\rm Li}$ with a neutron separation energy $S_n(^{11}{\rm Li}) = 0.396 \ {\rm MeV}$, which was suggested in \cite{Fornal:2018eol} as a candidate for the nuclear dark decay search. Such a process would lead to
\bea
^{11}{\rm Li} \to \,^{10}{\rm Li}^* + \chi \to \,^9{\rm Li} + n +\chi \ .
\eea
Nevertheless, as pointed out in \cite{Pfutzner:2018ieu}, this signal would be difficult to separate from the $\beta$-delayed deuteron emission background.

It was also argued in \cite{Pfutzner:2018ieu} that the $^{11}{\rm Be}$ nucleus, characterized by a neutron separation energy of $S_n(^{11}{\rm Be}) = 0.502 \ {\rm MeV}$, is a much better candidate to look for nuclear dark decays. It has a halo neutron, which enables calculating the rate of $^{11}{\rm Be}$ dark decay without the knowledge of nuclear matrix elements. The main decay channels of $^{11}{\rm Be}$ are: $^{11}{\rm Be}\to \,^{11}{\rm B} + e+\bar\nu_e$ and $^{11}{\rm Be}\to\!\,^{11}{\rm B}^*\to  \,^{7}{\rm Li}+ \!\,^{4}{\rm He}+ e+\bar\nu_e$ with branching ratios $\sim 97.1\%$ and $\sim 2.9\%$, respectively. Apart from those, there is also a theoretically expected $\beta$-delayed proton emission  channel $^{11}{\rm Be}\to \,^{10}{\rm Be} + p$ with a branching ratio  $\sim 2 \times 10^{-8}$ \cite{Borge2013}.

Interestingly, in an experiment measuring the number of $^{10}{\rm Be}$ nuclei  from $^{11}{\rm Be}$ decays \cite{Riisager:2014gia} it was found that there were $\sim 400$ times more $^{10}{\rm Be}$ nuclei than expected based on the above branching ratio. It was proposed in \cite{Pfutzner:2018ieu} that this might be explained by the nuclear dark decays
\bea
^{11}{\rm Be} \to \,^{10}{\rm Be} + \chi + \phi \ ,
\eea
caused by the halo neutron undergoing the dark decay $n \to \chi\,\phi$, and it was demonstrated that  having ${\rm Br}(^{11}{\rm Be}\to \,^{10}{\rm Be} + \chi+\phi) \sim 8 \times 10^{-6}$ is consistent within Model 2. Further theoretical work  \cite{Ejiri:2018dun} specified that such a dark decay of $^{11}{\rm Be}$ is phenomenologically viable if $m_{\tilde\chi}> m_n - S_n = 939.064 \ {\rm MeV}$. The question was whether there are protons in the final state of $^{11}{\rm Be}$ decays. A positive answer would suggest the existence of a near-threshold resonance in $^{11}{\rm B}$ emitting protons, whereas lack of protons would point to a dark decay.

Three collaborations undertook the task of measuring this in four different experiments: one group at CERN--ISOLDE \cite{ISOLDE}, second group at the Isotope Separator and Accelerator (ISAC) at TRIUMF  \cite{Ayyad:2019kna} and at the  National
Superconducting Cyclotron Laboratory at Michigan State University \cite{Ayyad:2022zqw}, and third group at Florida State University \cite{Lopez-Saavedra:2022vxh}. The results of the first group have yet to be published. The results of the  ISAC--TRIUMF experiment  \cite{Ayyad:2019kna} indicate that the number of protons from $^{11}{\rm Be}$ decays roughly matches the number of $^{10}{\rm Be}$ nuclei found in  \cite{Riisager:2014gia}, thus ruling out neutron dark decays in the mass range $937.993 \ {\rm MeV} < M_f < 939.064 \ {\rm MeV}$ and suggesting that the large number of protons observed in \cite{Riisager:2014gia} was the result of a  near-threshold resonance in $^{11}{\rm B}$ that enhanced the rate of $\beta$-delayed proton emission in $^{11}{\rm Be}$ decays. The presence of this resonance was confirmed by the  NSCL--MSU experiment \cite{Ayyad:2022zqw}  and the FSU experiment \cite{Lopez-Saavedra:2022vxh}.
Such a resonance was also hinted by theoretical calculations \cite{Okolowicz:2019ifb,Nguyen:2022qzt}. 
Nevertheless, a reanalysis of the  \cite{Riisager:2014gia} data discarded the original result and led to an upper bound of ${\rm Br}(^{11}{\rm Be}\to \,^{10}{\rm Be} + {\rm anything}) \lesssim 2 \times 10^{-6}$ \cite{Riisager:2020glj}, in contradiction with the result of  \cite{Ayyad:2019kna}, indicating the need for further investigation.

{\center{
\subsection*{Beam and bottle experiments}}}

Since the neutron lifetime puzzle arises from the discrepancy  between two (1996 and 2013) beam results and multiple bottle results, its resolution relies to a large extent on the outcome of the presently operating two beam neutron lifetime experiments: at the  National Institute of Standards and Technology (BL2 experiment) \cite{DEWEY2009189,Hoogerheide:2019yfu} and  at the Japan Proton Accelerator Research Complex \cite{Nagakura:2017xmv,Nagakura:2019xul,Hirota:2020mrd,Sumi:2021svn}. There is also ongoing work on the improved version of the beam experiment within the BL3 collaboration at NIST \cite{Nelsen:2020lbr,2021APS..APRT11008F,2022APS..APRT12004W,JasonFry}. We also note that the NIST beam experiment measures protons, whereas J-PARC is sensitive to electrons from neutron $\beta$ decays, making the two experiments complementary.
The initial data from J-PARC reported in \cite{Hirota:2020mrd} are not yet sufficiently precise to favor any of the two results. 

Another direct way of probing the neutron lifetime discrepancy is to place a proton counter inside a bottle  detector \cite{Fornal:2018mhk}, which would enable measuring the decay rate to protons simultaneously with the total neutron decay rate, thus lowering the possibility of unaccounted for systematic errors. This new approach is currently being pursued within the Los Alamos UCN group under  the project UCNProBe  \cite{Tang,2021APS..DNP.QJ008H}. 

\vspace{6mm}

{\center{
\subsection*{Neutron lifetime from $\boldsymbol\beta$ decay parameters}}}

The neutron lifetime in the Standard Model is determined, through the relation in Eq.\,(\ref{nlife}), by the  $V_{ud}$ element of the CKM matrix and the ratio $\lambda$ of the axial-vector to  vector current coefficients in the $\beta$ decay matrix element. The value of $V_{ud}$ is measured very accurately in superallowed $\beta$ decays \cite{Hardy:2014qxa}, and its average  adopted by the  PDG is $|V_{ud}| = 0.97373(11)(9)(27)$ \cite{Workman:2022ynf}. 
The most
recent precise measurement of $\lambda$  from the energy
spectrum of $\beta$ decay electrons provided by  PERKEO III \cite{Markisch:2018ndu} is
\bea
|\lambda_{\rm Perkeo\,III}| = 1.27641\pm 0.00045 \pm 0.00033 \ .
\eea
However, the value of $\lambda$ determined from the energy spectrum
of protons measured in the aSPECT experiment \cite{Beck:2019xye} is significantly lower, 
\bea
|\lambda_{\rm aSPECT}| =  1.2677 \pm 0.0028 \ . 
\eea
There is also a more recent result from the aCORN experiment \cite{Hassan:2020hrj}  extracting $\lambda$ from the spectrum of electrons, but it has a sizable error,  $|\lambda_{\rm aCORN}| = 1.2796(62)$. For 
 more details regarding the $\beta$ decay parameters 
 see \cite{Dubbers:2021wqv}.

Figure \ref{fig:6} displays the parameter space $|\lambda|$ versus $|V_{ud}|$ with the PERKEO III result (brown) and the aSPECT result (orange) overlaid with the bands corresponding to the neutron lifetime beam average (green) from Eq.\,(\ref{beamm}),  bottle average (blue) from Eq.\,(\ref{bottleb}), and $|V_{ud}|$ determined from superallowed transitions. The PERKEO III result agrees with the bottle average, but the aSPECT experiment is consistent with the beam average, favoring the neutron dark
decay proposal as a solution to the neutron lifetime puzzle.

\begin{figure}[t!]
\centerline{\includegraphics[trim={2cm 1.0cm 1.7cm 0.2cm},clip,width=9.75cm]{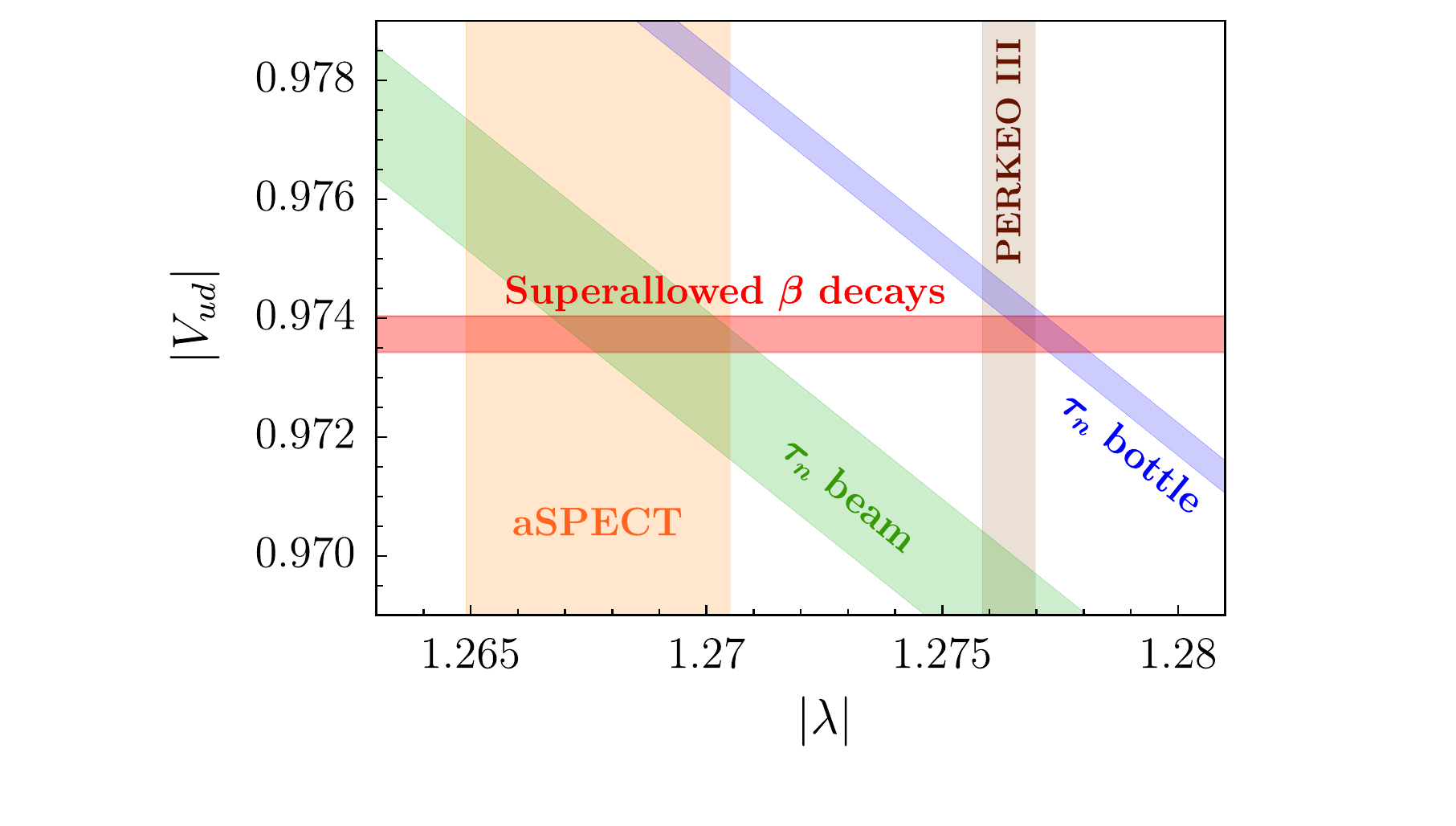}}
\vspace*{-4pt}
\caption{\small{Compilation plot of the $|\lambda|$ vs.\,$|V_{ud}|$ parameter space regions favored by the neutron lifetime beam average (green), bottle average (blue), superallowed $\beta$ decays \cite{Hardy:2014qxa,Workman:2022ynf} (red), PERKEO III  \cite{Markisch:2018ndu} (brown), and aSPECT  \cite{Beck:2019xye} (orange).}}\label{fig:6}
\vspace{2mm}
\end{figure}

\vspace{5mm}

{\center{
\subsection*{Other experiments}}}

Another way to determine  neutron lifetime is to measure neutrons from cosmic ray spallation of planets' and moons' surfaces (and atmospheres). Measurements of this type were performed using the neutron spectrometers onboard two of NASA's spacecraft missions: MESSENGER \cite{Wilson:2020sjk} which gave $\tau_n =780\pm60\pm70 \ \rm s$, and Lunar Prospector  \cite{Wilson:2020lfr} yielding $\tau_n =887\pm14^{+7}_{-3} \ \rm s$. 
Although those numbers are not competitive with beam and bottle results, future missions should be able to improve the accuracy to $\sim 1 \ \rm s$.  

Finally, one can also look for indirect signs of neutron dark decay at colliders, namely for the heavy color triplet scalar $\Phi$ mediating the decay. If in Model 1 the couplings satisfy $|\lambda_q\lambda_\chi |\lesssim10^{-4}$ or in Model 2 one has $|\lambda_q\lambda_{\tilde\chi}\lambda_\phi| \lesssim 10^{-4}$, then $\Phi$ would be accessible  at the Large Hadron Collider. 

\vspace{8mm}

\section{Theoretical developments} \label{s5}

The neutron dark decay proposal in \cite{Fornal:2018eol}  initiated not just experimental, but also intense theoretical investigations.
Those included studying neutron dark decay implications for neutron stars, which revealed the necessity of extending Models 1 and 2 to comply with the observed neutron star masses. It was also shown that certain models with neutron dark decay are capable of explaining both  the nature of dark matter and the origin of the matter-antimatter asymmetry of the Universe. Apart from that, novel experimental signatures of neutron dark decay models were proposed, arising from processes such as dark matter-neutron annihilation inside atoms or dark matter capture by atomic nuclei. Connections to other anomalies in particle physics were investigated as well.
Finally, it was pointed out that not only the neutron, but also other hadrons can undergo dark decays, with interesting predictions for hyperon, charm, and $B$ factories. We discuss those and other follow-up theoretical ideas  below.

{\center{
\subsection*{Neutron star masses}}}

Neutron stars do not become unstable due to the presence of a dark decay channel for the neutron since, similarly to the standard Pauli blocking of neutron $\beta$ decays by the degeneracy pressure of the protons and electrons, neutron dark decays are blocked by the degeneracy pressure caused by $\chi$ particles.  Nevertheless, a neutron dark decay channel would soften the neutron star equation of state and result in smaller than observed neutron star masses  \cite{Baym:2018ljz,McKeen:2018xwc,Motta:2018rxp}. In particular, within the framework of Models 1 and 2 the masses of neutron stars would not exceed $0.8 \, M_\odot$, which is significantly  below the $2\,M_\odot$ value for some of the observed neutron stars.

To solve this problem,  the neutron star equation of state has to be made stiffer, which  can be done by supplementing Models 1 and 2 with additional ingredients. In particular, neutron stars with $2\,M_\odot$ can exist in the presence of a neutron dark decay channel if  strong repulsive self-interactions in the dark sector are introduced \cite{Baym:2018ljz,McKeen:2018xwc,Motta:2018rxp} or repulsive  interactions between the  dark matter particle and the neutron \cite{Grinstein:2018ptl}. 
  Interestingly, such  self-interactions were proposed in the past to solve the $\Lambda$CDM cosmological model's small-scale structure problems  \cite{Spergel:1999mh}.

{\center{
\subsection*{Self-interactions in the dark sector} }}

Dark sector self-interactions  can be introduced by adding a dark vector gauge boson. A model of this type was constructed in  \cite{Cline:2018ami}, where the dark gauge boson is a dark photon  $A'$ and the neutron dark decay channel is $n\to \chi\,A'$. The Lagrangian of Model 1 is supplemented  with the terms
\bea
\mathcal{L}'_1 \ \supset \ -\tfrac14 F'_{\mu\nu}F^{\prime {\mu\nu}} - \tfrac12 \,\delta \,F_{\mu\nu}F^{\prime {\mu\nu}} \ ,
\eea
and the covariant derivative is modified to $D_\mu -i g'A'_\mu$, where $g'$ is the gauge coupling. The interaction  between $\chi$ and $A'$ is governed by $g'$ and produces repulsion between the $\chi$ particles. It was demonstrated in \cite{Cline:2018ami} that this model simultaneously accommodates ${\rm Br}(n\to \chi\,A') = 1\%$ and neutron stars with $2\,M_\odot$ for a wide range of  $\delta$ and $g'$ values. The particle $\chi$ can be a dark matter candidate, however, if thermally produced, it can account for only a small fraction of the dark matter in the Universe. This model can also be slightly extended to  explain the matter-antimatter asymmetry of the Universe through low-scale baryogenesis  \cite{Bringmann:2018sbs}.

It is also possible to build extensions of Model 2 which are consistent with all astrophysical constraints. One of such theories was constructed in  \cite{Karananas:2018goc} by augmenting the Model 2 Lagrangian  with terms including a dark gauge boson  $Z_D$, thus resulting in the following dark sector self-interactions,
\bea
\mathcal{L}'_2 \ \supset \ g' \bar\chi \,\slashed{Z}_{\!D}\chi - i\,g' \left(\phi^* \partial_\mu \phi-\phi\,\partial_\mu \phi^*\right)Z^\mu_D \ .
\eea
It was shown in   \cite{Karananas:2018goc}    that  one can have  ${\rm Br}(n\to \chi\,\phi) = 1\%$, the particle  $\chi$ (if non-thermally produced) can account for all of the dark matter in the Universe, and the self-interactions, making the model consistent with the observed neutron star masses, also provide a solution to the missing satellite problem, the ``too big to fail'' problem, and the
core vs. cusp problem.
\vspace{5mm}

{\center{
\subsection*{Dark matter-neutron repulsion}}}

An alternative idea allowing for consistency with neutron star constraints is to postulate additional repulsive interactions between the dark matter $\chi$ and the neutron \cite{Grinstein:2018ptl},  achieved by extending the Lagrangian of Model 2 by
\bea
\mathcal{L}''_2 \ \supset \ \mu \,H^\dagger H\,\phi + g_\chi\,\bar\chi \,\chi\,\phi \ .
\eea
This introduces an effective coupling $g_n \bar{n}n\phi$ via the Higgs portal, modifying the neutron star equation of state and, despite of the presence of the neutron decay channel $n\to\chi\,\phi$, permitting neutron stars to have $2\,M_\odot$.

{\center{
\subsection*{Stability of hydrogen}}}

It was pointed out in  \cite{McKeen:2020zni,Berezhiani:2018eds} that in  the case of Model 1 with $\chi$ being the dark matter particle, i.e., $m_\chi < m_p+m_e = 938.783 \ {\rm MeV}$, hydrogen becomes unstable with respect to the following dark decay channel,
\bea\label{hydro}
{\rm H} \to \chi\, \nu_e \ .
\eea
The branching ratio for the radiative contribution to this process, ${\rm H} \to \chi\, \nu_e\gamma$, is constrained by Borexino data \cite{McKeen:2020zni}. This leads to an additional constraint on the neutron dark decay channel $n\to \chi\,\gamma$ corresponding to the purple-shaded region in Figure\,\ref{fig:5}, which is excluded at  a  high confidence level. This implies that the allowed range of dark matter masses with   ${\rm Br}(n\to\chi\,\gamma) > 0.5\%$ is narrowed down to $m_\chi \gtrsim 938.5 \ {\rm MeV}$.

\vspace{8mm}

{\center{
\subsection*{Dark matter nuclear capture}}}\label{capture}

A new dark matter detection strategy is possible in models exhibiting  the neutron dark decay channel $n \to \chi\,\gamma$. Since $B_\chi=1$, the particle $\chi$ can be captured by atomic nuclei \cite{Fornal:2020bzz} because of the $\chi$-neutron mixing. This process is most interesting when $\chi$ is a dark matter candidate, since then $\chi$ from the galactic halo can be captured by a nucleus in a detector. If the nucleus is $(Z,A)$, the dark matter capture forms an excited nucleus $(Z,A+1)^*$, which subsequently  de-excites by emitting one photon or a cascade of photons, 
\bea
\chi + (Z,A)\,  \to \, (Z,A+1)^* \,\to\, (Z,A+1)+\gamma_{\rm cascade} \ ,
\eea
with  energy related to the  dark matter mass through
\bea
E_{\rm cascade} = S_n - (m_n-m_\chi) \ .
\eea
Here $S_n$ is the neutron separation energy for the nucleus $(Z,A+1)$. As a result, the energy of this cascade differs from the energy of a standard cascade (caused by neutron capture) by the mass difference $(m_n-m_\chi)$. 
This signature  can be searched for in  dark matter direct detection experiments, e.g., PandaX \cite{Cao:2014jsa}, XENONnT \cite{XENON:2022ltv}, and LUX-Zeplin \cite{LZ:2019sgr}, as well as 
large volume neutrino experiments, such as the future Deep Underground Neutrino
Experiment (DUNE) \cite{Abi:2020wmh}. The analysis in  \cite{Fornal:2020bzz} reveals that the discovery prospects are encouraging.

\vspace{15mm}

{\center{
\subsection*{Dark matter-neutron annihilation}}}

The particle $\chi$ produced in neutron dark decay need not necessarily be the dark matter particle, but it can rather be the antiparticle of dark matter. This would imply that the dark matter  in the galactic halo carries baryon number $B_{\bar\chi}=-1$, leading to spectacular signatures in direct detection experiments \cite{Jin:2018moh,Keung:2019wpw} (see also more general work in \cite{Davoudiasl:2010am,Davoudiasl:2011fj}), since dark matter can annihilate with nucleons inside nuclei via the following processes,
\vspace{1mm}
\begin{itemize}
\item[$\bullet$] Model 1:  \ \ \ $\bar{\chi} + n \to \gamma + {\rm meson(s)}$, \vspace{-2mm}
\item[$\bullet$] Model 2:  \ \ \ $\bar{\chi} + n \to \phi + {\rm meson(s)}$,
\end{itemize}
\vspace{1mm}
\noindent
in which the final state mesons experience very different kinematics than in standard nucleon decay considerations, and could be discovered in experiments such as Super-Kamiokande \cite{Fukuda:2002uc} and DUNE  \cite{Abi:2020wmh}. It was shown  \cite{Jin:2018moh,Keung:2019wpw} that this scenario (when $\chi$ is the antiparticle of dark matter) is experimentally excluded in the case of Model 1, but there exists a large parameter range for which it remains viable in the case of Model 2.

\vspace{1mm}

{\center{
\subsection*{Hadron dark decays}}}

The intermediate particles in neutron dark decay do not have to couple exclusively to first generation quarks. Allowing for nonzero interactions  with quarks of different  flavors (in this case ${\rm Br}(n \to \chi \,\gamma) \lesssim 10^{-6}$ \cite{Fajfer:2020tqf} due to the stringent flavor constraints from kaon mixing, but the dark decay $n\to \chi\,\phi$ remains unconstrained) leads to the possibility of hadrons other than the neutron  undergoing dark decays. This idea was first applied to neutral kaons and $B$-mesons in \cite{Barducci:2018rlx}, and later extended to other heavy hadrons in \cite{Heeck:2020nbq}, leading to apparent baryon number violation signals searchable not only in Super-K and DUNE, but also in charm and $B$ factories.

If the color triplet scalar $\Phi$ in Models 1 and 2 couples also to strange quarks, this leads to dark decays of hyperons   $\Lambda$, $\Sigma$, and $\Xi$. This scenario was investigated in great detail in  \cite{Alonso-Alvarez:2021oaj} and, in the case of the $\Lambda$ baryon, gives rise to the following dark decay channels,
\bea
\Lambda \to \chi\,\phi \ , \ \ \ \ \Lambda \to \pi^0 \chi \ , \ \ \ \ \Lambda \to \chi\,\gamma \ .
\eea
Building up on novel calculations of matrix elements relevant for hyperon dark decays, the expected rates at hyperon factories, such as BESIII \cite{BESIII:2021slv} and LHCb, were determined. It was demonstrated that prospects for the discovery of such hyperon dark decays are promising.

\newpage

{\center{
\subsection*{Other theoretical progress}}}

In addition to constructing neutron dark decay Models 1, 2 \cite{Fornal:2018eol} and their extensions with extra $\rm U(1)$ interactions \cite{Grinstein:2018ptl,Cline:2018ami,Bringmann:2018sbs,Karananas:2018goc} included to satisfy the neutron star constraints, further  theoretical progress was made on the model-building side, some of which we discuss below. 

It was shown in \cite{Elahi:2020urr} that the Standard Model can be extended by an ${\rm SU}(2)$ dark group, with the neutron dark decaying via $n\to \chi\,W'$ and $\chi$ constituting  all of the thermally produced dark matter in the Universe. It was suggested in \cite{Berezhiani:2018udo} that the neutron lifetime discrepancy is a result of neutron-mirror neutron oscillations; however, as was pointed out  in 
 \cite{Fornal:2019eiw}, to make this model consistent with experiment an extreme breaking of the $Z_2$ symmetry between the Standard Model and its mirror copy is necessary. A somewhat related proposal in 
 \cite{Tan:2019mrj} states that in the neutron-mirror neutron oscillation model the neutron dark decay can be mediated by nonperturbative effects. Finally, in \cite{Strumia:2021ybk} it was shown that there could exist neutron decay channels to three dark matter particles, $n\to \chi\chi\chi$, compatible with all bounds, including those from neutron stars, confirmed later by a more detailed analysis \cite{Husain:2022brl}.

In   \cite{McKeen:2020vpf} it was proposed that Model 1 with an extra dark fermion $\psi$ can explain the excess of  electron recoil events recorded by XENON1T  \cite{Aprile:2020tmw}, but that  anomaly disappeared  with more data collected by the XENONnT detector \cite{XENON:2022ltv}. 
In  \cite{TienDu:2020wks} novel neutron dark decay channels were proposed involving an intermediate new boson, which could be as light as $17 \ {\rm MeV}$ and provide a solution also to the $^8{\rm Be}$ nuclear transitions anomaly \cite{Krasznahorkay:2015iga,Feng:2016jff}.

Alternative  explanations of the neutron lifetime discrepancy include:\break  neutron-mirror neutron oscillations intensified in the presence of a magnetic field in beam experiments \cite{Berezhiani:2018eds} (experimentally excluded in \cite{Broussard:2021eyr}), a large Fierz interference term \cite{Ivanov:2018vit}, and quantum Zeno effect  \cite{Giacosa:2019nbz}. The suggestion was also put forward that in beam experiments there is an unaccounted for loss of protons due to their  collisions with the residual gas molecules in the quasi-Penning trap or other processes
\cite{Byrne:2019dhj,Serebrov:2020rvv,Byrne:2022lvp}. 

Finally, models of neutron dark decay have recently been shown to fit within the framework of asymmetric dark matter. This becomes possible if there are two or more types of dark matter particles and one of them has a mass below that of the neutron. Intriguingly, such models can be probed with future gravitational wave experiments \cite{Fornal:2023hri}.

\newpage

\section{Conclusions}\label{s6}

The neutron is one of the basic constituents of matter, indispensable for the existence of complex atomic nuclei. Despite being studied  throughout the last ninety years, the precise value of its lifetime is still an open question, with beam and bottle experiments providing different results. Knowing the value of the neutron lifetime is important since  it serves as an  input for Big Bang nucleosynthesis calculations, directly  affecting the primordial helium abundance. It also provides a clean test of the unitarity of the CKM matrix uninfluenced by nuclear structure effects. 

It was demonstrated that the neutron lifetime discrepancy between beam and bottle experiments can be explained if the neutron exhibits  a beyond-Standard-Model decay channel with a branching ratio $1\%$. Concrete particle physics models accommodating such a neutron dark decay channel were constructed and shown to be consistent with all  experiments and observations. Many of those models also provide  answers to other outstanding questions in particle physics, e.g., concerning the nature of dark matter and the origin of the matter-antimatter asymmetry of the Universe.

The neutron dark decay  creates a portal between the visible sector and the dark sector. As demonstrated in the original proposal and the  follow-up literature, it predicts novel signatures in various experiments across many fields: nuclear physics   (UCN, nuclear decays), dark matter direct detection (PandaX, XENONnT, LUX-Zeplin), neutrinos (Super-K, DUNE), colliders (LHC), and hyperon, charm, and $B$ meson  factories (Belle II, BESIII, LHCb). 
If any of the expected signals is discovered, pinning down the details of the dark decay channel will foster  a close collaboration between different groups,  bringing together   the particle and nuclear physics communities.

Finally, it is worth realizing that even if the beam and bottle results converge in the future, neutron dark decays with smaller branching ratios will still be an intriguing possibility to consider, enabling us to probe uncharted  parameter space of the dark matter sector.

\section*{Acknowledgments}

The author is very grateful to Benjam\'{i}n Grinstein for discussions and past collaboration. This work was supported by the National Science Foundation Grant No.\,${\rm PHY}$-$2213144$.

\bibliographystyle{utphys}
\bibliography{bibliography}

\end{document}